\documentclass[aps,superscriptaddress,twocolumn,noshowkeys,noshowpacs,nofootinbib]{revtex4-1}
\usepackage{epsf,amsmath,graphicx,amssymb,url,hyperref,dcolumn}
\usepackage{mathtools}
\usepackage{tabu}
\usepackage{color}
\usepackage[normalem]{ulem}
\usepackage{adjustbox}
\usepackage{array}
\usepackage{xcolor}
\usepackage[justification=raggedright,singlelinecheck=true]{caption}

\usepackage{subcaption}
\begin{document}
 \title{\bf Matter Dipole and Hubble Tension due to Large Wavelength Perturbations}
\author{Gopal Kashyap}
\email{gplkumar87@gmail.com}
\affiliation{Department of Physics, School of Advanced Sciences, Vellore Institute of Technology, Vellore 632014, Tamil Nadu, India}
\author{Naveen K. Singh}
\email{naveen.nkumars@gmail.com}
\affiliation{Sir P.T. Sarvajanik College of Science, Surat -395001, Gujarat, India}
\author{Pankaj Jain}
\email{pkjain@iitk.ac.in}
\affiliation{Department of Space, Planetary \& Astronomical Sciences \& Engineering (SPASE), Indian
Institute of Technology, Kanpur 208016, India
}
% \date{}

\begin{abstract}
We theoretically analyze the dipole anisotropy observed in the quasar distribution from the CatWISE2020 catalog. The catalog data shows a peak around $z\approx 1$, suggesting the presence of a large-scale dipole component. We explore the possibility that this dipole could be driven by primordial density fluctuations from modes that were superhorizon at the time of CMB decoupling but have since entered the horizon and become subhorizon. In particular, we consider the impact of adiabatic modes with wavenumbers $k$ in the range $(10^{-4} - 4 \times 10^{-3})~\mathrm{Mpc}^{-1} $, corresponding to wavelength scales of several Gpc. Such modes can create large-scale density variations, likely causing anisotropies in the distribution of matter and, as a result, affecting the number density of observed quasars. We also demonstrate that a superhorizon curvature perturbations mode, with comoving wavenumber $k\lesssim0.3H_0$ can lead to a significant enhancement in the locally inferred Hubble constant. This effect offers a viable explanation for the observed discrepancy between local and CMB inferred measurements of $H_0$.
\end{abstract}
\maketitle
\section{Introduction} \label{sec:Intro}
The standard cosmological model, $\Lambda$CDM, provides a remarkably successful framework for describing the large-scale structure and evolution of the universe. It is built on the cosmological principle, which is based on the assumption of statistical homogeneity and isotropy. The Universe is assumed to acquire this property during inflation \cite{Wald_1983} with small initial fluctuations growing through gravitational instability to form the cosmic web of galaxies we observe today. The cosmic microwave background (CMB) anisotropies provide strong support for this model, exhibiting a nearly scale-invariant spectrum of primordial fluctuations. The observed dipole in CMB is conventionally attributed to the motion of the Solar System with respect to the CMB rest frame \cite{Kogut1993,Hinshaw2009,fixsen1996,planck2014,planck2020_I,planck2020_II}. This relative motion can also induce a dipole in large-scale galaxy surveys as a result of Doppler and aberration effects \cite{ellis_1984}. This dipole was first extracted in \cite{Blake_2002} using the NVSS catalog \cite{1998AJ....115.1693C}. It was found that both the direction and magnitude are consistent with the CMB dipole. However, subsequent analysis using NVSS as well as other data sets found that although its direction is approximately consistent with the CMB dipole, its amplitude exceeds theoretical expectations \cite{Singal_2011, 10.1111/j.1365-2966.2012.22032.x, Rubart, TIWARI20151, Bengaly_2018, Secrest_2021,Secrest_2022}. For example, the dipole observed in quasar sources in the infrared CatWISE2020 catalog \cite{Marocco_2021} shows a 4.9 sigma deviation from the expected dipole \cite{Secrest_2021,Secrest_2022}. The reason for this discrepancy is currently unknown and may suggest the presence of additional physical effects beyond standard kinematic contributions. Furthermore, more data may be required to establish the signal. In Ref. \cite{Abghari_2024}, the authors suggested that more observations that are free from systematic effects are needed to declare a strong claim regarding the consistency of the CMB dipole with the dipole in quasars. In Ref. \cite{Cheng_2024}, it is concluded that NVSS dipole is consistent with a kinematic origin for the CMB dipole within $\Lambda$CDM. Further in Ref. \cite{Darling_2022}, it is shown that sky distribution and brightness of extragalactic radio sources are consistent with the amplitude and direction of dipole in CMB. In contrast, in Ref. \cite{Guandalin_2023}, inconsistency in the kinematic dipoles of the cosmic microwave background and of the large-scale matter distribution is discussed.

Interestingly, there also exist several additional effects that suggest a violation of statistical isotropy and homogeneity. These include dipole in radio polarizations \cite{Jain1998}, alignment of optical \cite{Hutsemekers_1998,Hutsemekers_Lamy_2001} and radio polarizations  \cite{Tiwari_Jain_2013,Pelgrims_Hutsemekers_2015,Tiwari_Pankaj_2016}, and radio jet axes \cite{Taylor,Panwar}, alignment of CMB quadrupole and octopole \cite{Quadrupole_and_octopole_alignment,Octo_Quadrupole_align}, hemispherical anisotropy in CMB  \cite{Eriksen_2004_hemispherical_power_asymmetry}, large scale bulk flow observations \cite{Kashlinsky_2008} and anisotropy in the Hubble constant \cite{PhysRevD.105.103510}. These violations of the cosmological principle are nicely reviewed in Ref. \cite{Kumar}. We emphasize that all these are indicative of a potential violation but not conclusive.

In this work, we investigate the contribution of large wavelength perturbations to the observed dipole anisotropy in galaxy number counts and their potential impact on the Hubble tension. Specifically, we consider the modes that were super-horizon at the time of CMB decoupling but have since entered the horizon and become sub-horizon in order to study the dipole. Previous studies have primarily explored the impact of super-horizon fluctuations on large-scale anisotropies \cite{Erickcek:2008,Erickek1_2008,Das_2021,Domenech_2022,Tiwari_2022}. These modes, extending beyond the observable universe, have been proposed as a possible source of the observed dipole. However, their contribution has been found to be negligible in affecting the dipole anisotropy in number counts \cite{Domenech_2022}. Superhorizon modes have been invoked \cite{Erickek1_2008} to explain the hemispherical anisotropy \cite{Eriksen_2004_hemispherical_power_asymmetry}. Although single-field models are unable to explain the observed amplitude, a superhorizon perturbation in the curvaton model of inflation \cite{PhysRevD.42.313} is consistent with all observations, including the CMB dipole, quadrupole, and octopole \cite{Erickek1_2008}.
We also consider the contribution of super-horizon modes on the Hubble constant. The contribution of sub-horizon modes on this parameter is found to be negligible. 

In a recent analysis of the CatWISE2020 data \cite{Mohit_2024}, it has been found that the dipole in number counts shows a sudden change as a function of color. The sources in the range $0.8<W1-W2<1.1$, referred as low color bin, show a dipole closely aligned with the CMB dipole. In contrast, sources in the color bin $1.1<W1-W2<1.4$, referred as high color bin, show a completely different behaviour with the dipole pointing roughly opposite to the galactic center. Here $W1$ and $W2$ refer to the infrared bands centered at wavelengths 3.4 $\mu$m, and 4.6 $\mu$m, respectively. Sources in both color bins satisfy the standard mid-infrared "red" criterion, $W1 - W2 \ge 0.8$. The lower color bin sources peak at a redshift of approximately $z\sim 0.88$, while the higher color bin sources peak at $z>1$. The reason for the change in behaviour is unknown and may arise from a galactic effect or be of cosmological origin. We speculate that the dipole at low color gets a significant contribution from sub-horizon modes. This is potentially interesting since such modes do give a non-negligible contribution to the dipole. In contrast, the super-horizon modes give negligible contribution and, hence, may explain the absence of cosmological dipole in the higher color bin.

Our basic assumption is that beyond a certain length scale $L_c$, there exist small deviations from the cosmological principle. Within the standard model, the modes corresponding to different wave numbers would be uncorrelated and point in different directions. Hence, their collective contribution is expected to cancel out and lead to a relatively small dipole on cosmological length scales. Here we assume that
modes beyond the length scale $L_c$ are coupled to each other and, in particular, may point in the same direction. This coupling may arise in some models which go beyond the standard Big Bang model. It has been suggested that even within the Big Bang model, large wave length modes may not respect the principle of statistical isotropy and homogeneity \cite{Rath:2013bfa}. Here we do not go into a detailed model and assume that beyond a certain distance scale all the modes point in the same direction and behave as the adiabatic modes in $\Lambda CDM$ model. We determine their collective effect by first assuming that it can be represented by a single mode of a certain wavelength. We next incoherently add the contributions of all modes beyond a certain wavelength.
Due to the assumed coupling between these modes, they can, in principle, act very differently from the perturbation modes in $\Lambda CDM$ model. In particular, they can add coherently to each other, leading to significant enhancement of amplitude compared to a single mode. However, in the present paper we do not pursue such models.

The large wavelength sub-horizon modes can contribute to the matter dipole, and we determine whether their contribution can explain the dipole seen in the low-color bin. We do not attempt to explain the behaviour seen in the high color bin, which we assume arises due to some unknown contribution from our galaxy. This is possible since the dipole in this bin points roughly opposite to the galactic center. However, it is not possible to rule out the possibility that this dipole is also of cosmological origin and may require a generalization of our proposal. Furthermore, it is possible that the dipole in both color bins may arise due to some unknown bias. This can only be settled by more refined data.

The very large wavelength superhorizon modes do not contribute to the matter dipole but can contribute to the local Hubble constant measurements because of their effect on the luminosity distance and observed redshift. If these perturbations substantially affect the inferred expansion rate at low redshifts, they could offer a possible explanation for the discrepancy between local and global $ H_0 $ measurements, known as Hubble tension \cite{Verde2019}.

This paper is organized as follows. In Sec. \ref{review}, we review the covariant formalism used to compute galaxy number counts. In Sec. \ref{number_count}, we analyze the impact of subhorizon perturbations on the number count dipole and provide an estimate of the resulting dipole amplitude. In Sec. \ref{hubble_tension}, we examine how superhorizon monopole perturbations can influence the locally measured value of the Hubble constant. Finally, we present our conclusions in Sec. \ref{conclusion}.

\section{A Review of Covariant Formulation of Galaxy Number Counts}\label{review}
In this section, we outline the covariant framework for estimating fluctuations in relativistic galaxy number counts, following the formalism of previous works \cite{Ellis2009, Kasai1987, Domenech_2022}. Consider sources with number density $ \hat{n}_s $ and 4-velocity $ \hat{u}^{\mu}_s $ in a perturbed flat FLRW universe with metric $ \hat{g}_{\mu \nu} $. Assuming $ \hat{n}_s $ depends only on time, photons emitted by these sources carry 4-momentum $ \hat{k}^{\mu} $ and travel along null geodesics parameterized by an affine parameter $ \hat{\lambda} $, reaching the observer at point $ o $ from direction $ \hat{n} $.  

The observed number of sources within distance $ d\hat{l}_s $ is given by  
\begin{eqnarray}  
d N(\hat \lambda, \hat n) = \hat{n}_s d\hat{l}_s d\hat{S}_s.  
\end{eqnarray}  
Here, $ \hat{n}_s $ is the number density at point $ s $, and photons reaching the observer at $ o $ travel along null geodesics. The infinitesimal distance $ d\hat{l}_s $ relates to $ d\hat{\lambda} $ as  
\begin{equation}  
d \hat{l}_s = \hat{\omega} d\hat{\lambda}, \quad \text{where} \quad \hat{\omega} = - \hat{k}_{\mu} \hat{u}_s^{\mu}  
\end{equation}  
is the photon energy. Since the observer's 4-velocity satisfies $ u_o^{\mu} \ll c $, it does not significantly affect the observed cross-sectional area $ d\hat{S}_s $. However, its variation with respect to $ \hat{\lambda} $ is given by  
\begin{equation}  
\frac{d (d \hat{S}_s)}{d \hat{\lambda}} = d \hat{S}_s \hat{\nabla}_{\mu} \hat{k}^{\mu}.  
\end{equation}  
One may write cross sectional area as $d\hat S_s = \hat d_o^2 d \Omega_o$, where $\hat d_o$ is observed area distance and $d \Omega_o$ is solid angle of the photon bundle at the observer position. To simplify the calculations, one can apply the following transformations \cite{Ellis2009}, which makes the galaxy number count invariant and also satisfy the Raychaudhuri equation for null geodesics,
\begin{eqnarray}
 && \hat g_{\mu \nu} = a^2 g_{\mu \nu} ,\,  d \hat \lambda = - a^2 d \lambda , \ \ \hat k^{\mu} = - a^{-2} k^{\mu}, \ \ \hat u^{\mu} = a^{-1} u^{\mu}, \nonumber \\
 && \hat d_0 = a d_o,  \hat n_s = a^{-3} n_s,   \hat \omega =- a^{-1} \omega =  a^{-1} k_{\mu} u_s^{\mu}
\end{eqnarray}
In these new variables those are without hat, galaxy number count can be written as,
\begin{eqnarray}
 d N (\lambda, \hat n) = n_s \omega d_o^2 d \lambda d \Omega_o.
\end{eqnarray}

Now, to calculate the number count fluctuations resulting from cosmological perturbations, start with defining the all perturbed variables with a tilde. So, in a perturbed flat FLRW metric in conformal coordinates $(\eta, x^i)$, the perturbed metric $\tilde{g}_{\mu \nu}$ is given by,
\begin{equation}
\tilde{g}_{\mu \nu} = \eta_{\mu \nu} + \delta g_{\mu \nu},
\end{equation}
where
\begin{equation}
\delta g_{\mu \nu} = -2\Psi \delta^{\mu}_{\nu} + 2\Phi \delta_{ij}.
\end{equation}
The 4-velocities of the source and observer, respectively, are given by:
\begin{equation}
\tilde{u}^\mu_{s/o} = (1 - \Psi_{s/o}) \delta_0^\mu + v_{s/o}^i \delta_i^\mu.
\end{equation}

Similarly, the other variables can be written as,
\begin{eqnarray}
\tilde{k}^{\mu} &=& k^{\mu} + \delta k^{\mu}, \\
\tilde{\omega} &=& 1 + \delta \omega, \\
\tilde{d}_o &=& d_o (1 + \delta r), \\
\tilde{n}_s &=& n_s (1 + \delta n) \\
\tilde{\eta} &=& \eta + \delta \eta.
\end{eqnarray}
The expression for $\delta k^{\mu}$, $\delta r$, $\delta \omega$, and $\delta \eta$ in terms of $\Phi$, $\Psi$, and $v_{s/o}$ are obtained as solutions of the Raychaudhuri equation. Following the detailed procedure outlined in \cite{Domenech_2022}, the first-order solutions to Raychaudhuri equation take the form,
\begin{align}
\label{eq:deltaomega}
\delta\omega_s &=\Psi_o - \Psi_s + n_iv_s^i-n_iv_o^i - \int_0^{\lambda_s} d\lambda_1\frac{\partial}{\partial \eta}(\Psi - \Phi)\,,\\
\label{eq:deltaeta}
\delta\eta_s &= \lambda_s(n_iv^i_o - \Psi_o) + 2\int_{0}^{\lambda_s}d\lambda_1\Psi \nonumber \\
&\hspace{1.5cm} +\int_0^{\lambda_s}d\lambda_1\int_0^{\lambda_1}d\lambda_2\frac{\partial}{\partial \eta}(\Psi - \Phi)\,,\\
\label{eq:deltalambda}
\delta\lambda_s &= \delta\eta_s-\frac{\delta\omega_s}{{\cal H}_s}\,,\\
\label{eq:deltar}
\delta_{r,s} &= \Phi_o+\int_0^{\lambda_s}\frac{d\lambda_1}{r^2(\lambda_1)}\int_0^{\lambda_1}d\lambda_2 \,r^2(\lambda_2) \nonumber\\ & \times\left\{\frac{\partial^2}{\partial\eta^2}\Phi - 2\frac{\partial}{\partial\eta}\frac{\partial}{\partial r}\Phi - \frac{1}{2}\Delta(\Psi - \Phi) + \frac{1}{2}\frac{\partial^2}{\partial r^2}(\Psi + \Phi)\right\}\,,
\end{align}

Furthermore, since our observations are based on the measured redshift $\tilde{z}$, the relationship between the background redshift $z$ and the observed redshift $\tilde{z}$ can be established by introducing a shift $\delta\lambda_s$ in the affine parameter $\lambda_s$ of the source, which arises due to cosmological perturbations. This shift is defined as,
\begin{align}
\label{definitionlambdas}
1 + \tilde z(\lambda_s + \delta\lambda_s) \equiv 1 + z(\lambda_s).
\end{align}
Expanding around $\lambda_s$ we get,
\begin{equation}
\tilde z(\lambda_s)=  z(\lambda_s) - {\cal H}_s(1+z(\lambda_s))\delta \lambda_s.
\label{deltaz}
\end{equation}

Additionally, since galaxy surveys generally count galaxies with observed fluxes above a specified threshold $F_* $, the number density of galaxies can be defined accordingly as
\begin{align}
\label{eq:calNs}
\hat {\cal N}_s(z,F>F_*) = \int_{ L_s(z,F)}^\infty\, dL \, \hat n_s(z, L)\,,
\end{align}
where $L_s$ is the source luminosity given by
\begin{align}
\label{eq:L}
L_s=4\pi F \hat d_o^2 (1 + z)^4=4\pi F \tilde d_o^2 a^2(\tilde\eta)(1 + \tilde z)^4\,.
\end{align}
Using the expression for source luminosity given in Eq. (\ref{eq:L}), one can derive the fluctuations in the number count of galaxies per redshift bin and per unit solid angle as,
\begin{eqnarray}
 \triangle_{\cal N}(z_s, \hat n) = \triangle_n^{\delta} - 2 x(z) \triangle_n^x - f_{evo}(z) \triangle_n^{evo}, \label{Delta_corr}
\end{eqnarray}
where, $x(z) = \frac{\partial \ln {\cal N}_s}{\partial \ln L}$ is magnitude distribution index, $f_{evo}= - \frac{\partial \ln {\cal N}_s}{\partial \ln (1+z_s)}$ is evolution bias and,
\begin{align}
 \triangle_n^{\delta} &= b(z) \delta_{mc}+ 2 \frac{\delta \lambda_s}{\lambda_s} + 2 \delta_{r,s} + \frac{d \delta \lambda_s}{d \lambda_s} \nonumber \\
 &\hspace{3cm}+ 3 {\cal H} _s v_s + \delta \omega_s \\
\triangle_n^x  &= \delta \omega_s
 + \frac{\delta \lambda_s}{\lambda_s} + \delta_{r,s} \\
  \triangle_n^{evo} &= {\cal H}_s v_s - \delta \omega_s.
\end{align}
Here, $ \delta_{mc} $ represents the comoving cold dark matter fluctuation, which is linearly related to the comoving galaxy number count fluctuation $ \delta_{nc} $ through the bias factor $ b(z) $ that varies with redshift \cite{Durrer_2020},
\begin{equation}
     \delta_{nc} = b(z) \delta_{mc} 
\end{equation}

The detail derivation of Eq. (\ref{Delta_corr}) is given in Ref. \cite{Domenech_2022}. We can take different model for $b(z), x(z)$ and $f_{evo}$ in calculating $\triangle_{{\cal N}}(z_s,\hat n)$.

\section{Effect of Subhorizon Perturbations on Number Count Dipole} \label{number_count}
The Planck observations have provided a highly precise measurement of the CMB dipole amplitude, determined to be\cite{planck_2020},
\begin{equation}
D_1^{\text{CMB}} = (1.23357 \pm 0.00036) \times 10^{-3}.
\end{equation}
This measurement is an important reference in cosmology, as it helps determine how fast our Solar System is moving relative to the CMB. It also allows us to compare dipole patterns seen in other cosmic datasets, such as quasar distributions and radio sources. 
If we assume that the observed dipole is purely kinematic in origin (i.e., due to the motion of the Solar System relative to the Cosmic Rest Frame (CRF)), we obtain the following velocity and direction:
\begin{align}
v_{\odot} &= (369.82 \pm 0.11)~\text{km~s}^{-1}, \\
(l, b) &= (264.021^\circ \pm 0.011^{\circ},~ 48.253^\circ \pm 0.005^{\circ}) .
\end{align}
As described in the introduction, the dipole seen in galaxy surveys shows a dipole whose amplitude is larger than expected from kinematic effects \cite{1998MNRAS.297..545B,Blake_2002, Singal_2011, 10.1111/j.1365-2966.2012.22032.x, Rubart, TIWARI20151, Bengaly_2018, Secrest_2021}. The dipole amplitude and direction in quasar CatWISE2020 catalog in the low color bin, $0.8<W1-W2<1.1$, is found to be\cite{Mohit_2024},
\begin{equation}  
d_N = 0.016 \pm 0.002,  \quad (l, b) = (283^\circ \pm 11^\circ,~ 29^\circ \pm 8^\circ),
\end{equation}  
which is significantly larger than the dipole predicted from CMB observations and is roughly aligned with the CMB dipole direction ($\sim 2 \sigma$ uncertainity). For the high color bin, $1.1<W1-W2<1.4$, amplitude and direction of dipole are found to be,
\begin{equation}  
d_N = 0.030 \pm 0.003,  \quad (l, b) = (194^\circ \pm 7^\circ,~ 19^\circ \pm 4^\circ),
\end{equation}
pointing roughly opposite to the galactic center and corresponds to higher redshift sources ($z>1$). In this work we do not consider such sources.

If the observed number dipole in low color bin is caused purely by our motion relative to the CMB, the estimated solar velocity is
\begin{equation}
v_\odot \approx 900 \pm 113 , \text{km/s},
\end{equation}
which differs from the CMB dipole velocity with a significance of about 4.7$\sigma$.

In a previous study\cite{Domenech_2022}, it was demonstrated that adiabatic superhorizon modes do not contribute significantly to the dipole anisotropies observed in galaxy number counts. However, as these modes re-enter the horizon over cosmic time, they begin to interact with matter and can influence the distribution of large-scale structure. In this work, we investigate the impact of such modes, initially superhorizon at the time of CMB decoupling but now slightly subhorizon, on number count fluctuations observed at redshift $ z \sim 1 $.

We consider modes in the range $10^{-4} \leq k \leq 4 \times 10^{-3} \,\text{Mpc}^{-1}$, which satisfy $k \eta_{\text{dec}} \ll 1$, while $k \eta_0 > k \lambda_s \gg 1$. These are superhorizon at decoupling but subhorizon today. The amplitude of primordial scalar perturbations is constrained by the observed CMB angular power spectrum, particularly through the Sachs–Wolfe and Integrated Sachs–Wolfe (ISW) effects at large angular scales \cite{sach_2012,Hu_2002}. The amplitude parameter $A_s \approx 2.1 \times 10^{-9}$, as reported by the Planck collaboration \cite{Planck_2018_X}, implies that during matter domination $\Phi_i \sim 10^{-5}$. For the range of wavenumbers considered here, we therefore adopt a conservative upper limit $\Phi_i \leq 5 \times 10^{-5}$, which remains consistent with current CMB observational constraints. Modes larger than $5\times 10^{-3} \text{Mpc}^{-1}$ contribute predominantly to the clustering dipole \cite{Secrest_2021}. This approach allows us to interpret the observed dipole in number counts as arising from both the kinematic dipole, attributed to our local motion, and an intrinsic dipole component which may arise from primordial subhorizon fluctuations, i.e, 
\begin{equation}
   d_N^{\rm obs} = d_N^{\rm{kin}} + d_N^{\rm{int}}. 
\end{equation}
Therefore, to account for the observed dipole in the number count, the amplitude of the intrinsic dipole arising from primordial subhorizon fluctuations should be approximately
\begin{equation}
    d_N^{\rm{int}} \simeq 0.009 \pm 0.002.
\end{equation}
Here, we assume the observed dipole direction is consistent with the CMB dipole upto $\sim 2 \sigma$ significance.

For a single subhorizon mode, where $ kr > 1 $, the cosmological perturbations can be expressed as a series expansion in terms of spherical harmonics $ Y_{lm} $ and spherical Bessel functions $ j_l $. Assuming the subhorizon mode aligns with the number count dipole, we can write $ {\rm \Phi}(r,k)$ as,

\begin{equation}
{\rm \Phi}(r,k) = \Phi(k)e^{i \mathbf{k} \cdot \mathbf{r}} = \sum_{l=0}^{\infty}\Phi(k)\, i^l (2l + 1) j_l(kr) P_l(\cos \theta),
\label{bessel_eq}
\end{equation}
where $\Phi(k)$ is the Fourier component of the perturbation. A similar expansion can be applied to isocurvature, density, and velocity perturbations. This representation allows us to analyze the contribution of individual modes to large scale anisotropies, particularly in the context of dipole measurements. With adiabatic initial conditions, the time evolution of $\Phi_l $ is governed by the linear equations for cosmological perturbations in Fourier space, expressed in conformal Newtonian gauge as \cite{MUKHANOV1992203,Domenech_2022},

\begin{equation}
    \Phi'' + 3\mathcal H(1 + c_s^2)\Phi' + \left(\mathcal H^2(1 + 3c_s^2) +2\mathcal H'\right)\Phi + c_s^2 k^2 \Phi=0,
    \label{Einstein_eq}
\end{equation}
where Hubble parameter is given by,
\begin{align}
{\cal H}={\cal H}_0\,a\sqrt{\Omega_{\Lambda,0}+\Omega_{{\rm m},0}a^{-3}+\Omega_{{\rm r},0}a^{-4}}\,,
\end{align}
and speed of sound, $c_s^2$, in Radiation+Matter universe is,
\begin{align}
c_s^2\equiv\frac{4}{9}\frac{\rho_{\rm r}}{\rho_{\rm m}+4\rho_{\rm r}/3}\,.
\end{align}

\begin{figure}[ht!]
\centering
\includegraphics[width=0.9\columnwidth]{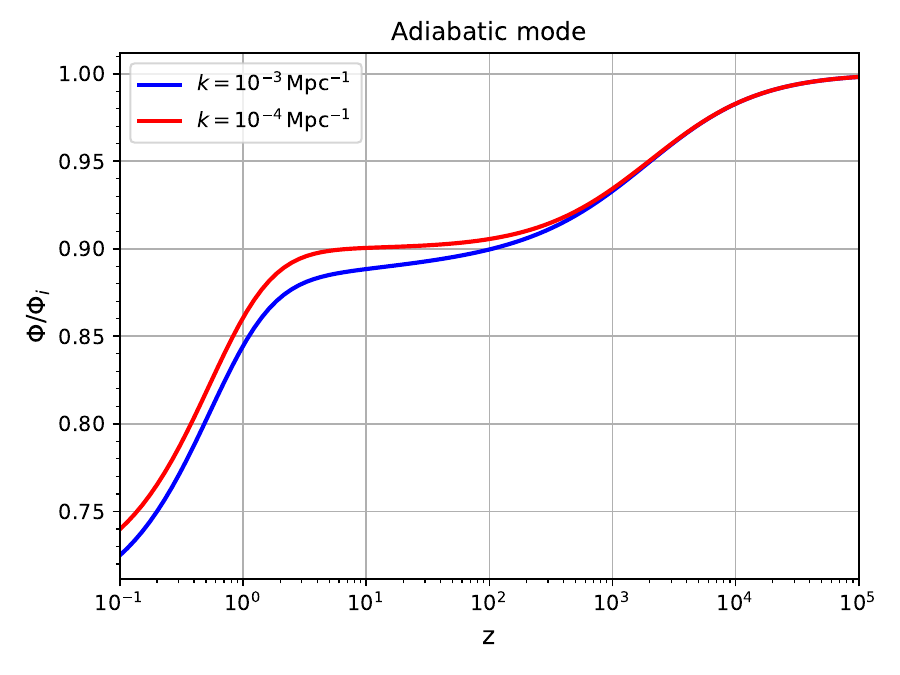}
\caption{The transfer function for $\Phi$ as a function of redshift, normalized to the initial curvature perturbation $\Phi_i$. By redshift $ z \sim 0.1 $, the initial perturbation has decayed by approximately 25\% of its original value, highlighting the impact of cosmic expansion and the influence of dark energy.}
\label{phi_evo}
\end{figure}

The Poisson equation relating the curvature perturbation $\Phi$ to the density perturbations is given by
\begin{equation}
 k^2 \Phi + 3 \mathcal{H} \left(\Phi' + \mathcal{H} \Phi \right) = 4 \pi G a^2 \left( \rho_m \delta_m + \rho_r \delta_r \right),
\end{equation}
where $\delta_m$ and $\delta_r$ denote the perturbations in cold dark matter and radiation, respectively. Using the solutions for $\Phi$, the evolution of velocity and matter density perturbations are given by \cite{Domenech_2022}
\begin{eqnarray}
v_m' + \mathcal{H} v_m - \Phi = 0,\\
\delta_m' - k^2 v_m = 0.
\end{eqnarray}
Here, the derivatives are with respect to the conformal time $\eta$.
We took the best-fit value of various cosmological parameters \cite{Planck_params} and numerically solve Eq. \ref{Einstein_eq}, setting the conditions during the radiation-dominated era to accurately track the evolution of perturbations from early times. Additionally, we incorporate the effect of the cosmological constant $\Lambda$, which leads to the further decay of perturbations at redshifts $ z < 1 $ due to the accelerated expansion of the universe. The resulting time evolution of $\Phi$ is shown in Fig. \ref{phi_evo}, showing the influence of both CDM and dark energy on the perturbation dynamics.
\subsection{Number Count Dipole} \label{rms_amp}
\begin{figure*}[htbp]
    \centering
    \begin{subfigure}[b]{0.48\textwidth}
        \centering
        \includegraphics[width=\textwidth]{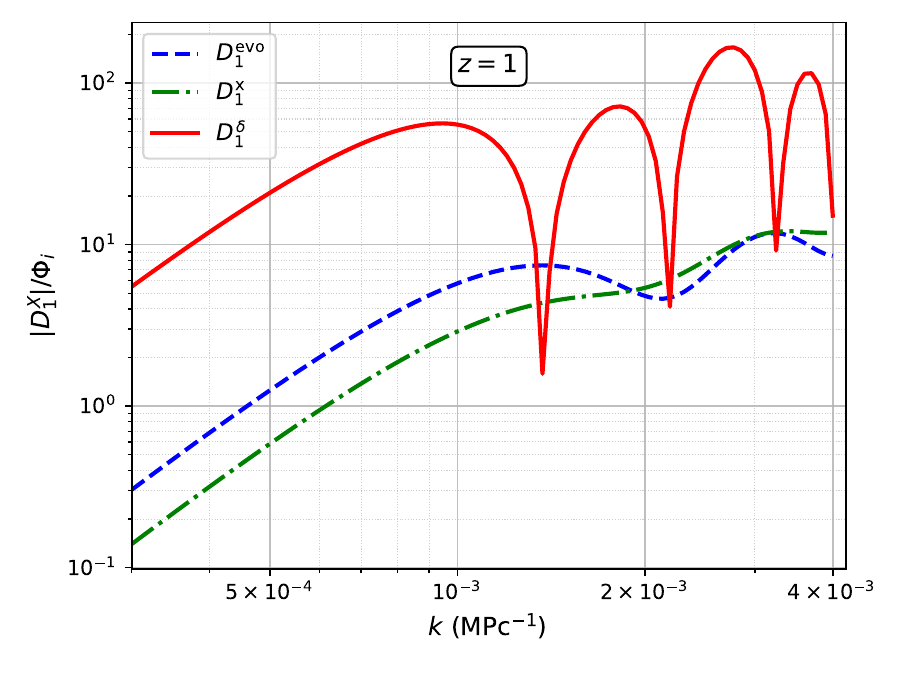}
        \caption{}
        \label{D1_k}
    \end{subfigure}%
    \hspace{0.04\textwidth}%
    \begin{subfigure}[b]{0.48\textwidth}
        \centering
        \includegraphics[width=\textwidth]{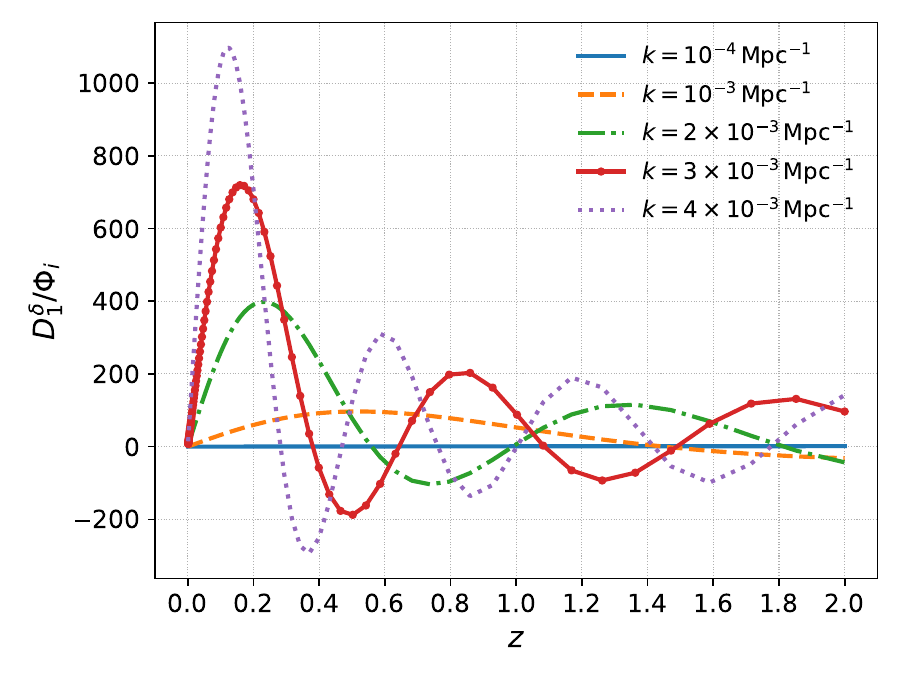}
        \caption{}
        \label{D1_z}
    \end{subfigure}
    \caption{(a) Contributions of various terms in Eq. \ref{Delta_corr} to the number count dipole at redshift $ z = 1 $ for wavenumber $ k $ in the range $ (10^{-4}, 4 \times 10^{-3}) $ Mpc$^{-1}$. (b) Contributions of $D_1^{\delta}$ to the number count dipole as a function of $z$.}
    \label{fig:combined}
\end{figure*}

To analyze the number count dipole, we focus on the dipole contribution ($ \ell = 1 $) of Eq. \ref{bessel_eq} to the number count dipole (\ref{Delta_corr}). In this case, the curvature perturbation $ {\rm \Phi}(r,k) $ becomes 
\begin{equation}    
{\rm \Phi}(r,k) = \Phi_1(k) \, 3i j_1(kr) \cos \theta,
\end{equation}
where $ j_1(kr) $ is the spherical Bessel function of order one, and $ \cos\theta $ accounts for the angular dependence of the dipole mode. We expand the number count fluctuations \ref{Delta_corr} similarly as,
\begin{equation} \label{D1_x}
    \triangle_n^X= D_1^X \cos\theta,
\end{equation}
where $X= (x,\delta, \rm evo)$. Using this expansion, we numerically analyze how the different contributions to the number count dipole in Eq. \ref{Delta_corr} depend on the wavenumber $ k $ for a pure adiabatic mode at redshift $ z = 1 $. Unlike the previous studies that assumed a constant bias factor ($ b = 1 $) \cite{Domenech_2022}, we incorporate a redshift dependent galaxy bias, given by \cite{Tiwari_2023}
\begin{equation}
    b(z) = 0.5 z^2 + 0.53z +1.54.
\end{equation}
This approach provides a more refined representation of galaxy clustering and its impact on dipole anisotropies at a given redshift. The computed contributions, plotted as a function of $ k $, are shown in Fig. \ref{D1_k}.

At redshift $ z = 1 $ and for the given range of $ k $-modes, Fig. \ref{D1_k} shows that the primary contribution to the number count dipole comes from the $ D_1^\delta $ term. In comparison, the contributions from other terms, such as $ D_1^x $ and $ D_1^{\rm evo} $, are significantly smaller and have a negligible impact on the total dipole amplitude. A precise determination of the dipole, however, would require a detailed understanding of the redshift evolution and magnitude distribution index of the sources, both of which remain uncertain. In Fig.~\ref{D1_z}, we also present the redshift evolution of $D_1^\delta$ for several $k$-modes. The plot shows that $D_1^\delta$ exhibits an oscillatory behavior with multiple peaks, among which the most prominent peak occurs at redshifts $z < 1$ for the considered $k$-range. Since the quasar number counts in the CatWISE sample peak around $z \sim 1$ \cite{Secrest_2021, Mohit_2024}, it is reasonable to assume that, at this redshift, the intrinsic dipole in the total number counts is primarily dominated by the matter density contrast contribution $D_1^\delta$. While this conclusion is specific to CatWISE catalogue, we note that other quasar catalogs peaking at different redshifts may exhibit different dipole contributions depending on their redshift distributions and selection functions \cite{WISE_2015,WISE_2018,Lyke_2020}.

\begin{figure}[ht]
\centering
\includegraphics[width=0.9\columnwidth]{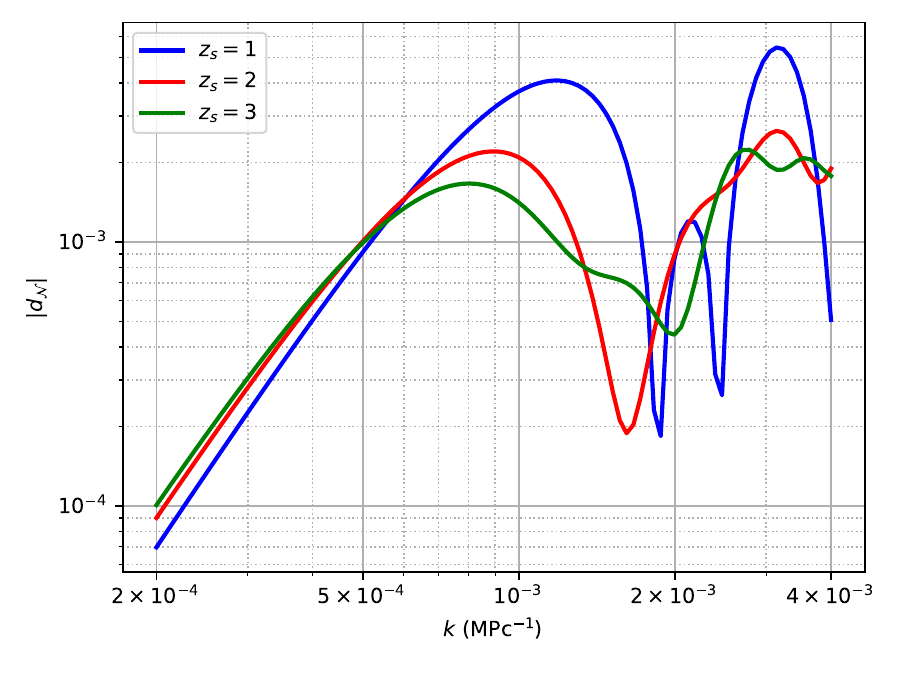}
\caption{The intrinsic dipole amplitude in the number counts is shown as a function of wavenumber $k$ for sources extending up to redshifts $ z_s = 1,2,$ and $3$. The curves are smoothed by interpolating the computed data points.}
\label{dN_intr}
\end{figure}

For a given $k$-mode, the intrinsic dipole in the total number count is given by \cite{Domenech_2022},
\begin{align} \label{dn_intrinsic}
d_{\mathcal{N}}^{\text{int}}(k) \approx \frac{1}{\int_0^{z_s} \mathcal{N}_s(z) r(z)^2 \frac{d\lambda}{dz}dz}
\int_0^{z_s} D_1^\delta(z,k) \mathcal{N}_s(z) r(z)^2 \frac{d\lambda}{dz}dz.
\end{align}
To compute the comoving number density $\mathcal{N}_s(z)$, we adopted the normalized redshift distribution function $P(z)$, fitted to the CatWISE quasar sample of lower color bin \cite{Mohit_2024}.
%as described in \cite{Domenech_2022},
\begin{align} \label{norm_pdf}
P(z) =& \left( \frac{1.12}{\sqrt{2\pi \Delta_1^2}} \right) \exp\left(- \frac{\log^2[z/0.48]}{2\Delta_1^2} \right)\\ \nonumber
&+ \left( \frac{0.28}{\sqrt{2\pi \Delta_2^2}} \right) \exp\left( -\frac{\log^2[z/0.88]}{2\Delta_2^2} \right),
\end{align}
where $\Delta_1=0.97$ and $\Delta_2=0.25$. This sample peaks at $z\sim0.88$.
We numerically compute the intrinsic dipole amplitude in the number counts, $d^{\text{int}}_{\mathcal{N}}(k)$, as defined in Eq. \ref{dn_intrinsic}, over the $k$ range $\left( 10^{-4},\ 4 \times 10^{-3} \right)\ \mathrm{Mpc}^{-1}$. For numerical evaluation, we discretize this range into 50 equally spaced modes in logarithmic scale, over which $d^{\text{int}}_{\mathcal{N}}(k)$ is computed. The redshift integration is performed from $z = 0$ up to the source redshift $z_s$, considering three cases: $z_s = 1,\ 2$, and $3$. The comoving number density of sources, $\mathcal{N}_s(z)$ is taken to be proportional to the normalized redshift probability distribution function, $P(z)$, as defined in Eq. \ref{norm_pdf}.  We do not consider $z_s>3$, as the source redshift distribution $P(z)$ peaks at $z \sim 0.88$ and becomes negligible beyond $z \sim 3$, making any additional contribution to the integrated dipole amplitude insignificant. Since $D_1^\delta$ is normalized to the initial curvature perturbation $\Phi_i$, we multiply the final expression for $d^{\text{int}}_{\mathcal{N}}(k)$ by $\Phi_i = 5 \times 10^{-5}$, consistent with the upper bounds on primordial curvature fluctuations from the CMB observational constraints. This allows us to estimate the maximal expected contribution of intrinsic large-scale fluctuations to the observed number count dipole. The resulting variation of $d^{\text{int}}_{\mathcal{N}}(k)$ as a function of $k$ for different source redshift cuts is shown in Figure \ref{dN_intr}. 

\subsection{Number count dipole in the standard model}
\begin{figure}[h]
\centering
\includegraphics[width=0.9\columnwidth]{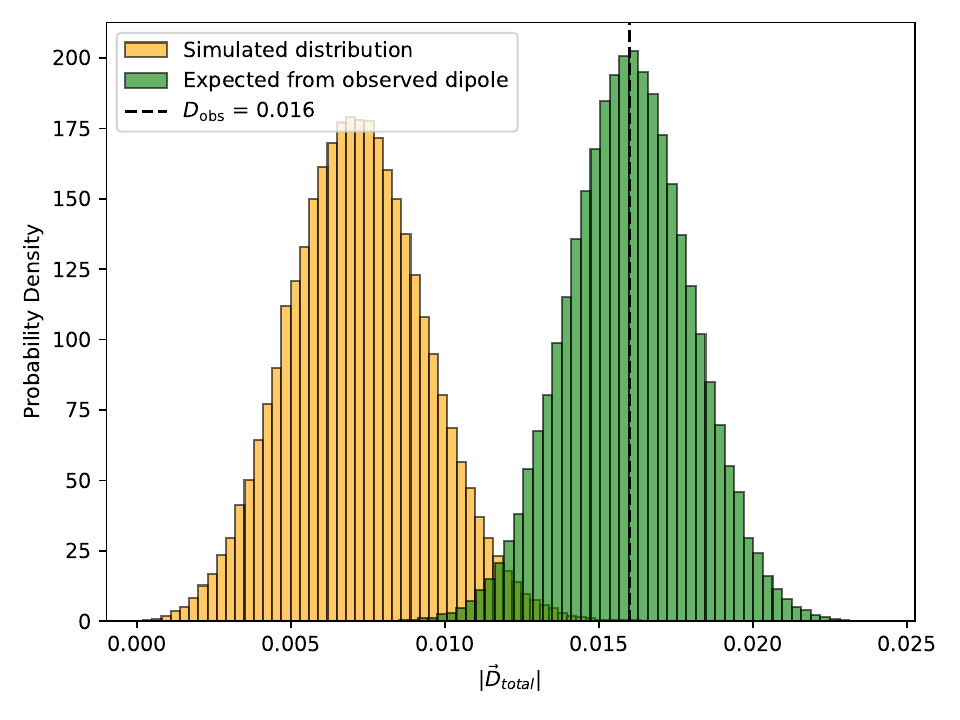}
\caption{Distribution of total dipole amplitude $|\vec{D}_{\rm total}|$ from random Gaussian samples, each combining an isotropic intrinsic dipole $\vec{D}_{\rm int}$ and the kinematic dipole $\vec{D}_{\rm kin}$ due to our motion. Result shown for source population resdshift cut at $z_s=1$.}
\label{D_total}
\end{figure}

In the standard cosmological model, perturbation modes corresponding to different wavenumbers are statistically uncorrelated and oriented in random directions due to isotropy. To model the dipole contribution from each mode, we represent the 3D intrinsic dipole vector $\vec{d}^{\rm int}_{\cal N}(k_i)$, for the case with a redshift cut at $z_s=1$, as a Gaussian random vector,
\begin{align}
&\vec{d}^{\rm int}_{\cal N}(k_i) = \left(d_{{\cal N}x}, d_{{\cal N}y}, d_{{\cal N}z}\right),
\end{align}
with,
\begin{align}
\left(d_{{\cal N}x}, d_{{\cal N}y}, d_{{\cal N}z}\right) \sim N(0, \sigma(k_i))^2),
\end{align}
where the standard deviation $\sigma(k_i)$ is related to the amplitude of the dipole at that scale by
\begin{equation}
\sigma(k_i) = \frac{|\vec{d}^{\rm int}_{\cal N}(k_i)|}{\sqrt{3}}, \quad (i= 1,2...50).
\end{equation}
To compute the cumulative contribution of long-wavelength modes to the intrinsic number count dipole, we evaluate the variance of the dipole amplitude, defined as 
\begin{equation} \label{variance}
\langle D_{\rm int}^2 (r)\rangle = \int_{k_{\rm min}}^{k_{\rm max}}|\vec d_{\cal N}(k)|^2 \mathcal{P}_\Phi(k)\, \frac{dk}{k},
\end{equation}
which is consistent with the definition of the dimensionless power spectrum $ \mathcal{P}_\Phi(k)$ as the variance per logarithmic interval in $k$ \cite{Dmitriy,Durrer_2020}. The RMS amplitude of the intrinsic dipole is then given by 
\begin{equation}
D_{\rm int}(r) = \sqrt{\langle D_{\rm int}^2(r) \rangle}.
\end{equation}
Here $ |\vec d_{\cal N}(k)| = \sqrt{d_{{\cal N}x}^2 + d_{{\cal N}y}^2 + d_{{\cal N}z}^2} $ denotes the magnitude of the dipole contribution from each mode, normalized by the initial curvature perturbation $\Phi_i(k)$.
We take a constant value of the primordial power spectrum, $ \mathcal{P}_\Phi(k) = 2.45 \times 10^{-9} $, which is a good approximation over the wavenumber range $ [10^{-4}, 10^{-3}]~{\rm Mpc}^{-1} $. The integral is evaluated numerically over 50 logarithmically spaced k-modes within this range. Using this result, a single realization of the total intrinsic dipole vector, $\vec D_{int}$, is then generated as a 3D Gaussian vector with zero mean and dispersion $D_{\rm int}/\sqrt{3}$ per component, ensuring consistency with the computed RMS amplitude. To this intrinsic total dipole vector, we add the expected kinematic dipole vector arising from our peculiar motion relative to the cosmic rest frame, given by\cite{ellis_1984}
\begin{equation}
\vec{D}_{\rm kin} = \left( 2 + x(1 + \alpha) \right) \beta\, \hat{n},
\end{equation}
where $ \hat{n} $ is a unit vector pointing in the direction $ (l = 264.02^\circ,\, b = 48.25^\circ) $. For the lower-color bin of the CatWISE quasar sample, we take $ \langle x \rangle = 1.66 $, $ \langle \alpha \rangle = 0.86 $ \cite{Mohit_2024}, and $ \beta = 370.0/c $, which gives a kinematic dipole amplitude $|\vec{D}_{\rm kin}|=0.0063$.
The total dipole vector is then given by,
\begin{equation}
\vec{D}_{\rm total} = \vec{D}_{\rm int} + \vec{D}_{\rm kin}.
\end{equation}
with total dipole amplitude $D_{total}$.
We repeat this process for 100000 times to construct the expected distribution of total dipole amplitudes. The resulting distribution is shown in Fig.\ref{D_total}.
 We compare this distribution with the observed dipole amplitude in lower color bin, $ D_{\rm obs}=0.016 \pm0.002 $\cite{Mohit_2024}, accounting for its uncertainty by generating random samples of $D_{obs}$ from a Gaussian distribution. The p-value is then defined as the fraction of all simulated and observed sample pairs satisfying $ |\vec{D}_{\rm total}| \geq D_{\rm obs} $. We find that the p-value is $p \sim 0.0022$, corresponding to a statistical significance $ \sim 2.85\sigma $. For the redshift cuts at $z_s=2$ and $z_s=3$, we find that the significance is $3.75\sigma$ and $3.9\sigma$, respectively.
This implies that, under the assumption that the observed dipole arises solely from the expected kinematic dipole and Gaussian distributed long wavelength perturbations, the observed dipole amplitude is significantly larger than what would be expected from standard $\Lambda$CDM cosmology.

\subsection{Number count dipole due to a single mode}
We next assume that only a single mode, which points in the direction of the CMB dipole, contributes. As explained in section \ref{sec:Intro}, this acts as an effective model for a correlated contribution from different modes all assumed to be pointing towards the CMB dipole. With a redshift cut $z_s=1$, we obtain the maximum amplitude of $5.3 \times 10^{-3}$ corresponding to the mode of wave number $k \sim 0.003\ \mathrm{Mpc}^{-1}$. This is less than two sigma smaller in comparison to the value required to explain the observed dipole. For larger redshift cuts of $z_s=2$, and $z_s=3$, the amplitude is much smaller and is unable to explain the observations.

\subsection{Number count dipole due to multiple modes}
We next assess the cumulative effect of a large number of modes, all assumed to be aligned with the CMB dipole, but otherwise uncorrelated with one another. We compute the variance of the intrinsic dipole amplitude, as defined in Eq.~\ref{variance}
\begin{equation} \label{dn_rms}
\langle D_{\rm int}^2 (r)\rangle = \int_{k_{\rm min}}^{k_{\rm max}} (d_{\cal N}(k))^2 \mathcal{P}_\Phi(k)\, \frac{dk}{k},
\end{equation}
% where $\Delta^2(k) = 2.42\times10^{-9}$ is the initial curvature perturbation variance per $ln (k)$ \cite{Planck_2018_X}. 
where $d_{\cal N}(k)$ is the amplitude of intrinsic dipole for a given mode normalized by initial curvature perturbation. This quantity captures the incoherent superposition of all contributing modes beyond a certain wavelength and provides a measure of the expected dipole amplitude.
By performing numerical integration of Eq.(\ref{dn_rms}) in the given k-range and for redshift cut at $z_s=1$, we find that the rms amplitude of intrinsic dipole is
\begin{equation}
(D_{\rm int})_{rms} \sim 4.1 \times 10^{-3}.    
\end{equation}
This is a little smaller than the maximum value we obtain for a single mode and deviates from the observed value by about 2.5$\sigma$. Similarly for the source redshift cut at $z_s=2$ and $z_s=3$, we find the 
$(D_{\rm int})_{rms}$ is approximately 0.0025 and 0.0021, respectively. The dipole value may be enhanced if the modes add coherently over some range of $k$ values. However, we do not consider this possibility in the present paper. 

We also find that modes with $ k < 10^{-4}~\mathrm{Mpc}^{-1} $ contribute negligibly to the dipole amplitude due to their superhorizon nature and suppressed growth. Furthermore, reducing the upper bound of the $ k $ range results in a decrease in the rms dipole amplitude. This suggests that modes near the upper end of the selected range dominate the observed dipole. It highlights the important role of slightly subhorizon fluctuations in shaping the large-scale anisotropy in the quasar distribution.

\subsection{CMB dipole}
We also examine the impact of adiabatic perturbations on CMB temperature fluctuations, which is given by \cite{Domenech_2022} as,
\begin{equation}
 \Delta T_{\mathrm{CMB}}(z_{\mathrm{dec}}, \hat{n}) = \frac{1}{4} \delta_{rc} + \mathcal{H}_{\mathrm{dec}} v_{\mathrm{dec}} - \delta \omega_{\mathrm{dec}}.
\end{equation}
Following the same procedure used for computing the number count dipole, we numerically evaluate the dipole in the CMB temperature fluctuations sourced by adiabatic modes. For illustration, we show the result for a representative mode with $k = 1 \times 10^{-3} \,\mathrm{Mpc}^{-1}$, for which we find a dipole amplitude of
\begin{equation}
\Delta T_{\mathrm{CMB}}(z_{\mathrm{dec}}) \simeq -0.8\, \Phi_i,
\end{equation}
For typical values of $\Phi_i \sim 10^{-5}$, this results in a dipole amplitude $\sim 10^{-5}$, which is significantly smaller than the observed CMB dipole ($\sim 10^{-3}$). We have verified that other modes within the range $k = 10^{-4}$ to $4.3 \times 10^{-3}\mathrm{Mpc}^{-1}$ yield similar amplitudes. This confirms that the dipole contribution from adiabatic modes in this range remains at the level of $10^{-5}$, and thus does not contribute significantly to the observed CMB dipole.

\section{Hubble Tension} \label{hubble_tension}
In the previous section, we explored how slightly subhorizon dipole perturbations affect anisotropies in the galaxy distribution. An equally important aspect is the influence of large wave length primordial curvature fluctuations on the locally inferred value of the Hubble constant \cite{Tiwari_2022}. While the dipole component of superhorizon-scale perturbations induces negligible anisotropy in the Hubble parameter \cite{Domenech_2022}, the monopole component, corresponding to a large-scale fluctuation, can cause a systematic shift in its local measurement. In contrast, subhorizon modes ($k \ge H_0$) contribute negligibly, as their small-scale inhomogeneities and velocity fields average out over large cosmological volumes. 

We consider the monopole component of the primordial curvature perturbation, as given in Eq.~\ref{bessel_eq}, incorporating a phase factor $\omega$,
\begin{equation}
\Phi(r,k) = \Phi_k\, j_0(kr)\, \sin \omega,
\label{super_mono}
\end{equation}
where $\Phi_k$ denotes the primordial amplitude of the superhorizon mode. The time evolution of this mode is governed by Eq.~\ref{Einstein_eq}.
To analyze the impact of this superhorizon mode on cosmic expansion, we examine its effect on the luminosity distance and on the redshift. Luminosity distance is given by,
\begin{align}
d_L=\hat d_o (1+z)^2\,,
\end{align}
where $\hat d_o$ is the angular diameter distance. Following the covariant formalism discussed in\cite{Domenech_2022}, the fluctuations in the luminosity distance become,
\begin{align} \label{ld_guillem}
\frac{\delta d_L}{d_L} = \delta_{r,s}+\delta\omega_s + \frac{\delta\lambda_s}{\lambda_s} \,.
\end{align}
Also, from Eq.(\ref{deltaz}), fluctuations in redshift is given by,
\begin{equation} 
    \frac{\delta z}{z} = - {\cal{H}}_s (\frac{1}{z}+1)\delta \lambda_s.
    \label{dz_z}
\end{equation}
At low redshifts, the fluctuation in the Hubble constant, arising from fluctuations in the luminosity distance and redshift, can be expressed as,
\begin{equation}  
\frac{\delta H_0}{H_0} \approx \frac{\delta z}{z}-\frac{\delta d_L}{d_L}.  
\end{equation}  
Using Eq. \ref{super_mono}, \ref{ld_guillem}, and \ref{dz_z}, the fluctuation in Hubble constant can be expressed as,
\begin{equation} 
\frac{\delta H_0}{H_0} = \Phi_k \sin \omega \,F(z,k).  
\label{delta_h0}
\end{equation}  
We numerically evaluate Eq.~(\ref{delta_h0}) for specified values of $\Phi_k$ and phase $\omega$ as a function of redshift $z$, and find that the resulting monopole contribution to $\delta H_0/H_0$ remains nearly constant across all wavenumbers $k$ within the superhorizon regime ($F(z,k)\sim F(z))$.

Local measurements of the Hubble constant, particularly those based on Cepheid-calibrated Type Ia supernovae~\cite{Riess_2022}, primarily rely on low-redshift sources ($z \lesssim 0.01$), making them sensitive to perturbations that can bias the inferred value of $H_0$. According to Ref.~\cite{Reid_2019}, the locally observed Hubble constant at $z = 0.0017$ is $H_0^{\text{obs}} = 73.5 \pm 1.4$ km\,s$^{-1}$\,Mpc$^{-1}$, while the global value inferred from \textit{Planck} data under the $\Lambda$CDM model is $H_0 = 67.36 \pm 0.54$ km\,s$^{-1}$\,Mpc$^{-1}$~\cite{Planck_params}. This discrepancy corresponds to a fractional difference $\delta H_0 / H_0 \simeq 0.0911$, reflecting a $\sim 9\%$ tension between local and global estimates of the Hubble constant.

To explain this excess, we set $\delta H_0/H_0 =0.0911$ in Eq. \ref{delta_h0} and evaluate the function F(z) at z=0.0017. We find that the required product of the primordial mode amplitude and phase factor is $\Phi_k\sin \omega =-1.1\times 10^{-4}$. Assuming initial phase equal to $3\pi/2$, this implies that a single superhorizon adiabatic curvature perturbation with amplitude $\Phi_k=1.1\times 10^{-4}$ is sufficient to produce the observed enhancement in the Hubble parameter.

\begin{figure}[h]
\centering
\includegraphics[width=0.9\columnwidth]{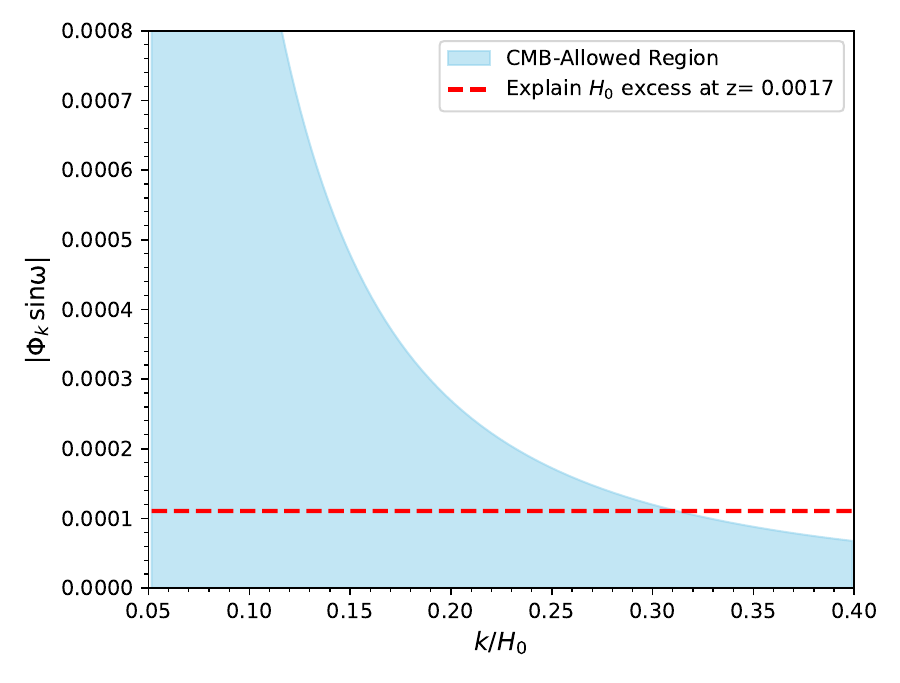}
\caption{ Plot of $|\Phi_k \sin\omega|$ versus $k/H_0$ for the observed
Hubble tension. The horizontal dashed line satisfies the $H_0$ excess at z=0.0017. The
blue-shaded region denotes the region of parameter space that
satisfies the CMB quadrupole constraint.}
\label{cmb_bound}
\end{figure}

 To further constrain the values of $\Phi_k$ and $\omega$, we impose bounds derived from CMB observations, following the analysis presented in Ref.~\cite{Erickcek:2008}. These constraints limit the possible values of the primordial superhorizon mode amplitude $\Phi_k$ and the phase $\omega$, ensuring that the perturbation remains consistent with the observed temperature anisotropies in the CMB. These bounds are given by 
 \begin{align}
    (k\, r_{dec})^2 |\Phi_k(z_{dec}) \sin \omega|&\leq 5.8 \cal Q \\
     (k\, r_{dec})^3 |\Phi_k(z_{dec}) \cos \omega|&\leq 32 \cal O,
 \end{align}
 where $\cal Q$ and $\cal O$  are 3 times the measured rms values of the CMB quadrupole
and octopole, respectively, and at decoupling, $\Phi_k(z_{dec}) = 0.937 \Phi_k$ \cite{Erickcek:2008}. We use the latest Planck values~\cite{Planck_params} of $Q \leq 1.69 \times 10^{-5}$ and $O \leq 2.44 \times 10^{-5}$. To account for the observed enhancement in the Hubble constant, the product 
 $\Phi_k\sin \omega$ required to be negative in our model. So we restrict the phase to the range $\pi < \omega < 2\pi$.   Since the CMB quadrupole provides more stringent constraints than the octopole, we consider only the quadrupole constraint when analyzing the parameter space. The corresponding upper limit on $|\Phi_k \sin \omega|$, derived from the CMB quadrupole constraint~\cite{Erickcek:2008}, is shown as the blue-shaded region in Fig.~\ref{cmb_bound}. For comparison, we also include the constraint from Ref.~\cite{Reid_2019}, which aims to account for the observed excess in the Hubble constant at $z = 0.0017$. 

As shown in Fig.~\ref{cmb_bound}, even a single superhorizon mode with wavenumber $k \lesssim 0.3 H_0$ can produce the necessary enhancement in the locally measured Hubble constant. This may provide a viable resolution to the observed Hubble tension.

\section{Conclusion} \label{conclusion}

In this study, we investigated the impact of large wavelength modes on both the dipole anisotropy observed in the quasar number counts and the local determination of the Hubble constant. 
The study is motivated by the recent observation that the cosmic dipole in the low color bin shows a very different behaviour from the high color bin \cite{Mohit_2024}. This essentially indicates a change in dipole with redshift, with the dipole at high redshift unlikely to have a cosmological origin. This is nicely consistent with the theoretical result that the predicted dipole amplitude decreases rapidly with an increase in the wavelength of the perturbation modes \cite{Domenech_2022}. 
We implement this by considering a single mode of sufficiently large wavelength. We also consider the case of several perturbation modes which are assumed to be pointing in the same direction beyond a certain wavelength, with their amplitudes uncorrelated with one another. Importantly, we do not provide a physical model for the origin of these modes, but simply assume their presence with amplitudes consistent with current observational constraints. Furthermore, we assume that they behave the same as the adiabatic modes in the $\Lambda CDM$ model.
Specifically, we considered the contributions of primordial superhorizon and slightly subhorizon modes to quantify their respective monopole and dipole effects on cosmological observables. In this work we have not considered any real model for the origin of such modes. In general, these modes may also contribute coherently over a certain range of wavelengths. Such generalizations are postponed to future study.

For dipole analysis, we focus on modes in the wavenumber $k$ range $(10^{-4}-4 \times 10^{-3})\,\mathrm{Mpc}^{-1}$, corresponding to scales just below the horizon at the redshifts relevant to the CATWISE2020 quasar sample in the color bin $0.8<W1-W2<1.1$ for which the observed dipole correlates with the CMB dipole \cite{Mohit_2024}. We find that a single mode with $k=0.003$ Mpc$^{-1}$ gives a substantial contribution but falls somewhat below the observed dipole. Our result is consistent with the earlier analysis of \cite{Domenech_2022}, with the distinction that we integrate Eq. \ref{dn_intrinsic} over sources up to redshift $z_s=1$ and incorporate a redshift-dependent galaxy bias $ b(z) $, whereas their analysis assumes a constant bias $ b = 1 $. Specifically, we use $ b(z) = 0.5 z^2 + 0.53z + 1.54 $, which exceeds unity even at low redshifts ($ z \lesssim 1 $). This leads to an enhancement in the estimated dipole amplitude by a factor of 2–3 relative to their findings. Furthermore, we extend our analysis to include source upto redshifts $z_s=2$ and $z_s=3$. We also assess the combined contributions of many modes, assuming that they all point in the same direction but added incoherently. By computing the root-mean-square dipole amplitude of the fluctuation in the total source count, we demonstrated that the incoherent superposition of these long-wavelength curvature perturbations contributes significantly to the observed dipole anisotropy in the galaxy distribution. However, their overall contribution is smaller than the maximum value obtained from a single mode. Hence, it falls significantly below the observed value. We point out that since the initial curvature perturbation for slightly subhorizon modes is taken to be consistent with the CMB constraint \cite{Planck_2018_X}, our result does not conflict with the observed CMB anisotropies. The matter dipole, along with its redshift dependence, is likely to be tested reliably at the Square Km Array (SKA) \cite{Ghosh:2023ayc}. If it confirms the higher dipole amplitude for $z<1$ and consistency with the CMB dipole at higher redshifts, it will provide strong evidence in favor of our proposed mechanism.

In addition, we have demonstrated that a single superhorizon adiabatic mode with wavenumber $ k \lesssim 0.3 H_0 $, consistent with CMB quadrupole constraints, can induce a monopole fluctuation in the local expansion rate. By evaluating the monopole contribution to the Hubble parameter at $ z = 0.0017 $, where the Cepheid-calibrated supernova data is calibrated, we find that a mode with amplitude $ \Phi_k \sim 1.1 \times 10^{-4} $ and phase $ \omega \sim 1.5\pi $ can explain the observed $\sim 9\%$ excess in $ H_0 $ relative to the global $\Lambda$CDM value. The resulting combination of $ \Phi_k \sin\omega $ lies well within the parameter space allowed by the latest Planck constraints on the CMB quadrupole moments. 

Together, these findings establish a framework in which both dipole anisotropies in galaxy distributions and shifts in the Hubble constant, may originate from a small violation of the cosmological principle at large distance scales. Although our calculated dipole amplitude falls below the observations, we emphasize that generalizations are possible which may be explored in future studies.

There exist many avenues for further tests of the large scale modes, discussed in this paper. As mentioned in the Introduction, a superhorizon mode within the curvaton model of inflation can also explain the hemispherical anisotropy in CMB \cite{Erickek1_2008}. Hence, it will be very interesting to explore the corrections to the Hubble constant within the curvaton model. We may also explore anisotropies in the Hubble parameter within this model. Although, a superhorizon mode does not produce a dipole in the Hubble parameter, it may produce higher modes, such as a quadrupole, which may be explored in future cosmological observations. The effect of the superhorizon mode may also be explored in other cosmological observables such as the epoch of reionization \cite{PhysRevLett.118.151301,Deshpande_2018} and baryon accoustic oscillations.

\section*{Acknowledgements}
GK acknowledges the Vellore Institute of Technology for providing financial support through its Seed Grant (No.SG20230035), year 2023.
\bibliographystyle{prsty}
 \bibliography{NVSS_dipole}
\end{document}